\documentclass{elsart}
\usepackage{graphicx}
\usepackage{amssymb}
\journal{Physics Letters A}
\begin{document}
\begin{frontmatter}
\title{Kraus decomposition for chaotic environments including time-dependent
subsystem Hamiltonians}

\author{Murat \c{C}etinba\c{s}\corauthref{cor}},
\corauth[cor]{Corresponding author.} 
\ead{cetinbas@sfu.ca}
\author{Joshua Wilkie}
\ead{wilkie@sfu.ca}
\address{Department of Chemistry, Simon Fraser University, Burnaby, British Columbia V5A 1S6, Canada}

\begin{abstract}
We derive an exact and explicit Kraus decomposition for the reduced density of a quantum system simultaneously interacting with time-dependent external fields and a chaotic environment of thermodynamic dimension. We test the accuracy of the Kraus decomposition against exact numerical results for a CNOT gate performed on two qubits of an $(N+2)$-qubit statically flawed isolated quantum computer. Here the $N$ idle qubits
comprise the finite environment. We obtain very good agreement even for small $N$. 
\end{abstract}

\end{frontmatter}

\section{Introduction}
A statically flawed isolated quantum computer (QC) \cite{QC,GS} can exhibit a rich variety of
dynamical behaviors\cite{CW1,CW2,CW4}, many of which are detrimental to its operation. Large unitary shifts have
been observed which can destroy the fidelity over the span of a single gate\cite{CW1,CW2,CW4}. Non-unitary effects
such as internal decoherence and dissipation can also affect performance\cite{CW2,CW4}. In some cases the
QC can act similar to a kicked top\cite{LEKT} in the exponential decay regime\cite{CW4}, in which case options for 
error correction
strategies\cite{QC,Brown} are quite limited. In other configurations, part of a QC can be employed to measure
the strength of residual two-body interactions\cite{CW1}, and this knowledge may prove useful for error correction.
Idle qubits can also be manipulated to improve performance of the active qubits\cite{CW2,CW4}. In fact
such self-interacting quantum systems have an exceedingly complex dynamics we are only
beginning to explore.

The statically flawed isolated QC can be viewed as a subsystem-bath model where the subsystem is the 
active part of the QC (i.e., a set of qubits which are manipulated with classical external fields)
and the bath consists of the idle qubits. The flaws consist of 
single body imperfections in qubits and residual two-body interactions. Similar subsystem-bath
models arise in many proposed quantum technologies which employ condensed phase architectures.
Quantum control of a chemical reaction in molecules\cite{SB,Rat} in a hosting medium (e.g. gas phase, solution, surface, 
solid), requires coherent manipulation of a few degrees of freedom (the reaction coordinates) while 
they simultaneously interact with the rest of the degrees of freedom of the molecule(s) and the 
degrees of freedom of the environment. The degrees of freedom of the rest of the molecule(s) are
usually strongly self-interacting, and hence are not well modeled by independent harmonic 
oscillators. These give rise to an intrinsic decoherence dynamics\cite{Brumer} like that in a QC. 
Similarly, the degrees of freedom of the hosting medium are generally
strongly self-interacting and anharmonic. These give rise to external decoherence. Thus, coherent quantum control in 
general consists of a coherently manipulated subsystem interacting with a strongly self-interacting multi-component
bath. 

At high temperatures there are a number of theories which can be employed to model such complex 
dynamics. Redfield theory \cite{Red,Opp} and its generalizations \cite{Coal,Neu} take into account internal 
bath dynamics via realistic molecular dynamics simulations. However, Redfield theory is notorious for 
its positivity violation, incorrect prediction of the long time limit, and for having divergent correlation 
functions for baths with discrete spectra. The theory of Bulgac et al\cite{Bulg} which represents the environment 
as an ensemble of random matrices is also useful for simulation of complex environmental dynamics. The semi-classical
Wigner method \cite{GB} can also prove quite accurate for moderately high temperatures. 

At very low temperatures the available computational schemes are more limited. An approximate master equation\cite{CW2,SRA1,SRA2,SRA3,CW5} which is a non-Markovian generalization of completely-positive-dynamical-semigroup theory\cite{dsg}, has been shown
to be accurate for some subsystems interacting with chaotic baths\cite{CW2,CW5}. A Kraus decomposition\cite{Kraus} for
independent subsystems interacting with chaotic thermodynamic baths has also been developed\cite{CW3}, and 
shown to be accurate numerically. Neither theory is exact, at least for finite baths, and so such methods 
are best used in tandem. Where they are in
agreement some confidence can presumably be placed in their predictions. 

Here we extend this Kraus decomposition to subsystems which also interact with classical external fields. This is
of course the most important case for coherent control problems.
We also consider a more general class of subsystem-bath interactions than was addressed in \cite{CW3}.
Finally, to obtain the greatest generality we make no assumptions regarding the Hermiticity of the 
subsystem Hamiltonian, so that the method can be employed with complex absorbing potentials\cite{Abs}.
The extended chaotic Kraus decomposition (CKD) we derive automatically satisfies all required conservation 
laws, i.e., Hermiticity, norm conservation (for Hermitian subsystems only) and complete positivity for the reduced density. It is only 
exact in the limit of a bath of thermodynamic dimension. We test the accuracy of our CKD against exact 
numerical results of a realistic isolated QC model and we obtain very good agreement.  

Our test model \cite{CW2,CW4} represents a two-qubit register performing a CNOT gate while interacting with 
neighboring idle qubits via static internal residual interactions. Statically flawed isolated QC models 
offer a prototypical example of a manipulated subsystem interacting with a self-interacting, possibly chaotic, 
environment wherein exact quantum dynamics can be readily obtained. These exact numerical results serve as 
a benchmark against which new theories of complex environments can be tested. Accordingly, we test the 
time-dependent extension of our CKD against these exact benchmark results \cite{CW2,CW4} for a large number 
of configurations: eight different initial states, two types of error generating interactions (bit-flip and phase errors),
 and a number of intra-bath couplings in the chaotic regime. We obtain very accurate results despite 
our relatively small bath dimension.

Organization of this manuscript is as follows: We present the time dependent extension of the CKD in section 2. In section 3 we discuss our isolated QC model. In sections 4 and 5 we test the Kraus decomposition against exact numerical results. In section 6 we discuss our results.

\section{Extension of chaotic Kraus decomposition}

We will now derive a CKD which can be applied more widely than that developed in \cite{CW3}. The general structure
of the argument is similar to that of \cite{CW3}, and where the two coincide we will refer the reader to 
the previous manuscript for details. We will begin by deriving an abstract general Kraus decomposition which
makes no assumptions about the properties of the Hamiltonians beyond that the total Hamiltonian is of the form
\begin{equation}
\label{Ham}
\hat{H}(t)=\hat{H}_{S}(t) + \sum_{\mu} \hat{S}_{\mu}\hat{B}_{\mu}+\hat{H}_{B}
\end{equation}
where $\hat{H}_{S}(t)=\hat{H}_{S}+\hat{\cal E}(t)$ consists of the time-independent native system Hamiltonian $\hat{H}_{S}$, and the time-dependent Hamiltonian representing external driving fields $\hat{\cal E}(t)$. $\hat{S}_{\mu}$ and $\hat{B}_\mu$ are interaction Hamiltonians in system and bath degrees of freedom, and $\hat{H}_{B}$ is a bath Hamiltonian. We will then use properties specific to chaotic thermodynamic baths to transform the Kraus decomposition
into a more tractable sum.
 
We consider a product initial condition 
\begin{equation}
\hat{\rho}(0)=\hat{\rho}_{S}(0)\otimes \hat{\rho}_{B}(0)
\end{equation} 
where $\hat{\rho}_{S}(0)$ is an arbitrary initial subsystem state, and $\hat{\rho}_{B}(0)$ is an 
initial bath state of canonical form, i.e.
\begin{equation}
\hat{\rho}_{B}(0)=\sum_j \frac{ e^{-E_{j}/k_BT }} {Q} | j\rangle \langle j |,
\end{equation}
where $Q=\sum_{k} \exp{ \{ -E_{k} / k_BT \}}$. Here $|j\rangle$ denotes a complete bath eigenbasis, i.e. $\hat{H}_{B}|k\rangle=E_{k}|k\rangle$ and $\sum_k|k\rangle\langle k|=\hat{{ I}}_B$. Obviously this is not the most general initial condition\cite{Copt}, nor is it the most likely
initial condition. However, the CKD requires such an uncorrelated initial state, and this is a distinct limitation of the method.
However, in many studies an uncorrelated initial state is assumed for simplicity or in ignorance of the true initial state, and
so the CKD may prove very useful in spite of this obvious limitation. Also, the CKD we derive can be applied to one type of initially
correlated state, e.g. $\hat{\rho}(0)=|\psi_1(0)\rangle\langle\psi_1(0)|\otimes \hat{\rho}_B^1(0)+|\psi_2(0)\rangle\langle\psi_2(0)|\otimes \hat{\rho}_B^2(0)$, where $\hat{\rho}_B^1(0)$ and $\hat{\rho}_B^2(0)$ are different thermodynamic chaotic baths of potentially differing temperatures. In this case, the CKD is applied independently to each term, and the sum is then an exact representation of $\hat{\rho}(t)$.
Finally, it is also worth pointing out that other master equation
based theories do not require such restrictions on the form of the initial states\cite{SRA1,SRA2,SRA3,CW5}.

\subsection{General Kraus decomposition}

Given these initial conditions and Hamiltonians the time evolved density is
$\hat{\rho}(t)=\hat{U}(t)\hat{\rho}(0)\hat{U}^{\dagger}(t)$
where the propagator $\hat{U}(t)$ is given by
\begin{equation}
\hat{U}(t)=\hat{\cal T}\exp{[-(i/\hbar)\int_0^t\hat{H}(t')dt']}
\end{equation}
and $\hat{\cal T}$ denotes the time ordering operator. 

The exact reduced density of the subsystem at time $t$, $\hat{\rho}_{S}(t)$, can be  expressed by as a partial trace over the bath degrees of freedom 
\begin{equation}
\hat{\rho}_{S}(t)={\rm Tr}_{B}\{ \hat{U}(t)\hat{\rho}(0)\hat{U}^{\dagger}(t) \}.
\end{equation}
Performing the partial trace operation in the bath eigenbasis gives
\begin{equation}\label{kr}
\hat{\rho}_{S}(t) = \sum_{j,k} \langle j| \hat{U}(t) \left( \hat{\rho}_{S}(0) \otimes \frac{ e^{-E_{k}/k_BT }} {Q} | k\rangle \langle k | \right) \hat{U}^{\dagger}(t) | j \rangle.
\end{equation}
Defining the Kraus operators $\hat{\cal K}_{j,k}(t)$ via
\begin{equation}\label{Kraus}
\hat{\cal K}_{j,k}(t)=\sqrt{p_{k}} \langle j| \hat{U}(t) | k\rangle,
\end{equation}
where $p_{k}=\exp{ \{ -E_{k} / k_BT \} } /Q$ are the initial populations of the bath eigenstates, it follows 
that Eq. (\ref{kr}) can be written in the Kraus decomposition form\cite{Kraus}
\begin{equation}\label{OSR}
\hat{\rho}_{S}(t)=\sum_{j,k} \hat{\cal K}_{j,k}(t) \hat{\rho}_{S}(0) \hat{\cal K}_{j,k}^{\dagger}(t).
\end{equation}
Note that if the subsystem Hamiltonian is Hermitian then it is always true that 
\begin{displaymath}
\sum_{j,k} \hat{\cal K}_{j,k}(t) \hat{\cal K}_{j,k}^{\dagger}(t)=\hat{I}_{S},  
\end{displaymath}
but Hamiltonians with absorbing potentials\cite{Abs} do not satisfy this relation. This is no way limits the usefulness of the 
decomposition.

\subsection{Chaotic Kraus decomposition}

We next employ properties of the coupling matrix elements of a chaotic thermodynamic bath to simplify
the forms of the Kraus operators $\hat{\cal K}_{j,k}(t)$ .
First we rewrite the total Hamiltonian (\ref{Ham}) in terms of the bath eigenvalues and eigenvectors, i.e.,
\begin{equation}
\label{Ham2}
\hat{H}=\hat{H}_{S}(t)+\sum_{\mu}\hat{S}_{\mu}\sum_{j,k}B_{\mu}^{j,k} | j \rangle \langle k |+\sum_{j}E_{j} | j
\rangle \langle j |,
\end{equation}
where $B_{\mu}^{j,k}=\langle j |\hat{B}_{\mu}|k\rangle$ are the matrix elements of the bath coupling operator in the
complete bath eigenbasis. 

Open system dynamics is largely governed by the properties of the coupling operators $\hat{S}_{\mu}$ and $\hat{B}_{\mu}$. 
If a bath is chaotic with $N$ degrees of freedom, then it can be argued\cite{CW3} that as a result of scaling the 
off-diagonal matrix elements of any operator $\hat{B}$, in the bath eigenbasis,  will vanish as $N\rightarrow \infty$
(i.e. $\langle j|\hat{B}|k\rangle \rightarrow 0$ for $j\ne k$). This result has been known for some time\cite{SemiC}, but it has an important application here. Applied to our bath coupling operators it means that
$B_{\mu}^{j,k}=0$ for $j\ne k$ in the limit of a thermodynamic bath.

Using this property, the total Hamiltonian (\ref{Ham2}) simplifies to  
\begin{equation}
\label{Ham3}
\hat{H}=\hat{H}_{S}(t)+\sum_k(\sum_{\mu}\hat{S}_{\mu}B_{\mu}^{k,k}+E_{k}) | k \rangle \langle k |.
\end{equation}
Taylor expansion of (\ref{Kraus}) using the Magnus formula\cite{Magnus} then gives
\begin{eqnarray}
&&\hat{\cal K}_{j,k}(t)=\sqrt{p_{k}}\langle j|[1+(-i/\hbar)\int_0^tdt_1~\hat{H}(t_1)\nonumber \\
&&+(-i/\hbar)^2\int_0^tdt_1\int_0^{t_1}dt_2 ~\hat{{\cal T}}\hat{H}(t_1)\hat{H}(t_2)\nonumber \\
&&+(-i/\hbar)^3\int_0^tdt_1\int_0^{t_1}dt_2\int_0^{t_2}dt_3 ~\hat{{\cal T}}\hat{H}(t_1)\hat{H}(t_2)\hat{H}(t_3)+\dots]|k\rangle
\end{eqnarray}
which, since $\hat{H}(t)$ is block diagonal in the bath eigenbasis, simplifies to
\begin{eqnarray}
&&\hat{\cal K}_{j,k}(t)=\sqrt{p_{k}}[1+(-i/\hbar)\int_0^tdt_1~\langle j|\hat{H}(t_1)|j\rangle\nonumber \\
&&+(-i/\hbar)^2\int_0^tdt_1\int_0^{t_1}dt_2 ~\hat{{\cal T}}\langle j|\hat{H}(t_1)|j\rangle\langle j|\hat{H}(t_2)|j\rangle \nonumber \\
&&+(-i/\hbar)^3\int_0^tdt_1\int_0^{t_1}dt_2\int_0^{t_2}dt_3 ~\hat{{\cal T}} \langle j|\hat{H}(t_1)|j\rangle \langle j|\hat{H}(t_2)|j\rangle \langle j|\hat{H}(t_3)|j\rangle\nonumber \\
&&+ \dots]\delta_{j,k}.
\end{eqnarray}
Now, using the Magnus Taylor expansion\cite{Magnus} in reverse gives
\begin{eqnarray}
\hat{{\cal K}}_{j,k}(t)=\sqrt{p_{k}}~\hat{{\cal T}}\exp \left\{ -\frac{i}{\hbar} \int_{0}^{t}dt^{\prime}(\hat{H}_{S}(t^{\prime})+\sum_{\mu}\hat{S}_{\mu}B_{\mu}^{k,k} + E_{k})  \right\} \delta_{j,k}\label{CKraus}
\end{eqnarray}
which in turn can be substituted into (\ref{OSR}) to obtain the final form of the CKD.
The double sum becomes a single sum and 
\begin{equation}\label{CTK}
\hat{\rho}_{S}(t)=\sum_{k} \hat{\cal K}_{k,k}(t) \hat{\rho}_{S}(0) \hat{\cal K}_{k,k}^{\dagger}(t)
\end{equation}
with $\hat{\cal K}_{k,k}(t)$ given by (\ref{CKraus}).

Eq. (\ref{CTK}) is our final result. It extends our previous CKD \cite{CW3} to time-dependent subsystem Hamiltonians, potentially 
non-Hermitian subsystem Hamiltonians, and more general subsystem-bath coupling operators. 

\subsection{Numerical strategy}

While Eq. (\ref{CTK}) is exact only for thermodynamic chaotic baths, our previous experience in \cite{CW3} 
suggests that it may be accurate even for quite small baths. Thus, it is important to address the issue
of how (\ref{CTK}) can be employed in practice. 

First, note that the sum over $k$ can probably be truncated at quite small values for very low temperature
systems like QCs. Furthermore, if $\hat{\rho}_{S}(0)=|\psi(0)\rangle\langle \psi(0)|$ then one can define states 
\begin{equation}
|\psi_k(t)\rangle=\hat{\cal T}\exp \left\{ -\frac{i}{\hbar} \int_{0}^{t}dt^{\prime}(\hat{H}_{S}(t^{\prime})+\sum_{\mu}\hat{S}_{\mu}B_{\mu}^{k,k} )  \right\}|\psi(0)\rangle
\end{equation}
such that
\begin{equation}
\hat{\rho}_S(t)=\sum_kp_k |\psi_k(t)\rangle\langle \psi_k(t)|\label{SumS}
\end{equation}
and 
\begin{equation}
d|\psi_k(t)\rangle/dt=-(i/\hbar)[\hat{H}_{S}(t)+\sum_{\mu}\hat{S}_{\mu}B_{\mu}^{k,k} ]|\psi_k(t)\rangle.\label{SEs}
\end{equation}
Now, equations (\ref{SEs}) can be solved exactly using standard Runge-Kutta\cite{RK} techniques.  In the case where
the initial state is not pure one could obviously find a similar set of Liouville-von Neumann equations,
\begin{eqnarray}
d\hat{\rho}_S^k(t)/dt=-(i/\hbar)[\hat{H}_{S}(t)+\sum_{\mu}\hat{S}_{\mu}B_{\mu}^{k,k}, \hat{\rho}_S^k(t)],
\end{eqnarray}
with $\hat{\rho}_S^k(0)=\hat{\rho}_S(0)$,
from which the reduced density can be constructed via $\hat{\rho}_S(t)=\sum_kp_k \hat{\rho}_S^k(t)$. Once again these equations can be 
solved numerically with standard techniques.
 
\section{Test model}

Our test model represents a two-qubit register performing a CNOT gate while interacting with idle neighboring qubits via static residual interactions. 
We consider a two dimensional circuit and so the number of idle qubits $N$ is ten, which clearly does not constitute a bath of thermodynamic 
dimension. It is however chaotic when the magnitude of the residual interactions is sufficiently large. This can be verified using the 
nearest neighbor eigenvalue spacing distribution\cite{Chaos,Chaos2} and using Loschmidt echo\cite{Peres,Emerson} calculations\cite{CW3}.
Note that we do not repeat these calculations here since the bath is the same in this and the previous study \cite{CW3}.
We will see that the CKD is accurate in the chaotic regime even though the bath dimension is small.

We consider the following total Hamiltonian for the $(N+2)$-qubit isolated QC 
\begin{equation} \label{HamQC}
\hat{H}(t)=\hat{H}_{S}(t)+\hat{H}_{SB}+\hat{H}_{B}.
\end{equation}
\noindent
Here $\hat{H}_{S}(t)$ represents the time-dependent control Hamiltonian that implements the CNOT gate, $\hat{H}_{B}$ is a bath Hamiltonian which represents the idle qubits, and $\hat{H}_{SB}$ is the interaction Hamiltonian which is responsible for the generation of errors.  

Elementary gate operations comprising the CNOT protocol are implemented using the control Hamiltonian \cite{Nori} 
\begin{equation}
\hat{H}_{S}(t)=-\frac{1}{2}\sum_{i=1}^2({\cal B}_{i}^{x}(t)\hat{\sigma}_{x}^{i} +{\cal B}_{i}^{z}(t)\hat{\sigma}_{z}^{i} )
+ {\cal J}_{x}(t)\hat{\sigma}_{x}^{1}\hat{\sigma}_{x}^{2}.
\label{ctrl}
\end{equation}
\noindent
In our CNOT implementation, we assume that the control Hamiltonian (\ref{ctrl}) is free of all imperfections, and that gates can be switched on and off by perfect square pulses. We also do not allow free system evolution. These conditions ensure a perfect implementation with maximum fidelity. 
Therefore, any error that is observed in the CNOT implementation, regardless of its type (i.e. unitary or non-unitary), is due to the subsystem-bath interactions. The subsystem elementary gate Hamiltonians and their corresponding time intervals can be found in Table 1.  

We consider two types of error generation, modeled by the interaction Hamiltonian
\begin{equation}
\hat{H}_{SB}= ( \hat{\sigma}_{\alpha}^{1}+\hat{\sigma}_{\alpha}^{2} )\hat{\Sigma}_{\alpha},
\end{equation}
where $\hat{\Sigma}_{\alpha}=\sum_{i=3}^{N+2} \lambda_{\alpha}^{i} \hat{\sigma}_{\alpha}^{i}$ with index $\alpha\in \{x,z\}$. 
Henceforth, we sometimes refer to the errors generated by $xx$-type coupling as bit-flip errors and the errors generated by $zz$-type coupling as phase errors.

\begin{table}
\caption{The switching intervals and active Hamiltonians used to implement CNOT gate.}
\begin{center}
\begin{tabular}{ccc}
\hline 
\hline 
Switching Intervals~~~~~~~&~~~Active Hamiltonian
\tabularnewline
\hline
$[\tau_{0}=0, \tau_{1} = \pi / (2{\cal{B}}^{z})]~~~~~$
&
$-\frac{1}{2}{\cal{B}}^{z}\hat{\sigma}_{z}^{2}$
\tabularnewline
$[\tau_{1}, \tau_{2} = \tau_{1}+ \pi / (2{\cal{B}}^{x})]~~~~$
&
$-\frac{1}{2}{\cal{B}}^{x} \hat{\sigma}_{x}^{2}$
\tabularnewline
$[\tau_{2}, \tau_{3} = \tau_{2}+ \pi / (2{\cal{B}}^{z})]~~~~$
&
$+\frac{1}{2}{\cal{B}}^{z}\hat{\sigma}_{z}^{2}$
\tabularnewline
$[\tau_{3}, \tau_{4} = \tau_{3}+ \sqrt{2} \pi / (2{\cal{B}}^{z})]$
&
$-\frac{1}{2}{\cal{B}}^{z}\sum_{i=1}^{2}(\hat{\sigma}_{z}^{i}+\hat{\sigma}_{x}^{i})$
\tabularnewline
$[\tau_{4}, \tau_{5} = \tau_{4}+ \pi / (4{\cal{J}}_x)]~~~~$
&
${\cal J}_{x}(-\hat{\sigma}_{x}^{1}-\hat{\sigma}_{x}^{2}+\hat{\sigma}_{x}^{1}\hat{\sigma}_{x}^{2})$
\tabularnewline
$[\tau_{5}, \tau_{6} = \tau_{5}+ \sqrt{2} \pi / (2{\cal{B}}^{z})]$
&
$+\frac{1}{2}{\cal{B}}^{z}\sum_{i=1}^{2}(\hat{\sigma}_{z}^{i}+\hat{\sigma}_{x}^{i})$
\tabularnewline
$[\tau_{6}, \tau_{7} =\tau_{6}+ \pi / (2{\cal{B}}^{z})]~~~~$
&
$-\frac{1}{2}{\cal{B}}^{z}\hat{\sigma}_{z}^{2}$
\tabularnewline
$[\tau_{7}, \tau_{8} =\tau_{7}+ \pi / (2{\cal{B}}^{x})]~~~~$
&
$+\frac{1}{2}{\cal{B}}^{x}\hat{\sigma}_{x}^{2}$
\tabularnewline
$[\tau_{8}, \tau_{9} =\tau_{8}+ \pi / (2{\cal{B}}^{z})]~~~~$
&
$+\frac{1}{2}{\cal{B}}^{z}\hat{\sigma}_{z}^{2}$
\tabularnewline
\hline\hline
\end{tabular}\end{center}
\end{table}

While we assumed a perfect control Hamiltonian we did not make the same assumption for the bath Hamiltonian. We let the bath Hamiltonian include 
one and two-body static internal flaws. It has been argued in a number of studies \cite{GS} that one and two-body internal flaws are unavoidable 
features of multi-qubit QCs. Our bath Hamiltonian is then an $N$-qubit generalization of the two-qubit control Hamiltonian (\ref{ctrl}) given by
\begin{eqnarray}
\hat{H}_{B} = -\frac{1}{2} \sum_{i=3}^{N+2} \left( B_{i}^{x}\hat{\sigma}_{x}^{i} 
+ B_{i}^{z} \hat{\sigma}_{z}^{i} \right)
+ \sum_{i=3}^{N+1}\sum_{j=i+1}^{N+2} J_{x}^{i,j} \hat{\sigma}_{x}^{i}\hat{\sigma}_{x}^{j}.
\label{Hb}
\end{eqnarray}
Here one-qubit parameters are sampled randomly and uniformly as $B_{i}^{\alpha} \in [B_{0}^{\alpha} \! -\delta/2, \: B_{0}^{\alpha} \! +\delta/2]$ where $B_{0}^{\alpha}$ is the average value of the distribution, and $\delta$ is a detuning parameter. Two-qubit residual interactions and system-bath interactions are also sampled randomly and uniformly as $J_{x}^{i,j} \in [-J_x,\ J_x]$ and $\lambda^{i}_{\alpha} \in [-\lambda,\ \lambda]$, respectively. 
We will consider a number of $J_x$ values to explore the transition to strong chaos.

We considered two sets of four initial system states $\hat{\rho}_{S}(0)=|\psi_{0}\rangle\langle\psi_{0}|$. The first set consists of four standard basis states 
\begin{equation}
|\psi_0\rangle \in \{|00\rangle,|01\rangle, |10\rangle, |11\rangle \},
\end{equation}
and the second set consists of four Bell states 
\begin{equation}
|\psi_0\rangle \in \{(|00\rangle \pm |11\rangle)/\sqrt{2}, (|01\rangle \pm |10\rangle)/\sqrt{2} \}.
\end{equation}

Thus, we have a set of benchmark tests which includes two types of subsystem-bath coupling, eight different initial states, and three $J_x$ values. In
total, this set of tests is thus quite stringent. Reporting all of this data would however be problematic. Fortunately, there is very little state specificity
in the data and so we can just report average quantities. That is, for each observable we average the results over the four standard initial
states and the four Bell initial states, separately.

\subsection{Quantitative error measures}

We quantified the extent of deviation from the ideal system evolution by two error measures: average purity and fidelity. Averages are taken over the four states of each set, i.e. over the four initial standard basis states, and over the four initial Bell states.  

The average purity is defined by
\begin{equation}
{\bar{\mathcal P}}(t)=\frac{1}{4}\sum_{|\psi_{0}\rangle}{\rm Tr}_S [\hat{\rho}_{S}^{2}(t) ]
\end{equation}
The purity measures the degree of deviation from unitary dynamics and thus it quantifies non-unitary errors such as decoherence and dissipation. 

The average fidelity is defined by 
\begin{equation}
{\bar{\mathcal F}}(t) = \frac{1}{4}\sum_{|\psi_{0}\rangle}{\rm Tr}_S [\hat{\rho}_{S}(t)\hat{\rho}_{S}^{ideal}(t)]
\end{equation}
where $\hat{\rho}_{S}^{ideal}(t)$ 
is the ideal subsystem evolution in the absence of interactions with the idle qubits. The fidelity measures how close the actual 
density stays to the ideal density in the course of the dynamics. Hence, it is sensitive to both unitary and non-unitary errors. 
Since the purity is insensitive to unitary effects, large deviations between purity and fidelity can be used as an indicator of unitary errors 
induced by the coherent shift process\cite{CW1,CW2,CW4}. 

The ideal value of the purity/fidelity is unity for pure initial conditions. 

We will also examine individual matrix elements, to see how the errors are manifested in the reduced density.

\subsection{Numerical parameters}

In our simulations we used the experimentally accessible control parameters for the charge-qubit QCs \cite{Nori,Makhlin} for which ${\cal{B}}^{\alpha}\!=1.00 \ {\epsilon}$, ${\cal{J}}_{x}=0.05 \ {\epsilon}$ and $kT=0.25\ {\epsilon}$, in units of $\epsilon=200$ mK. We assumed that the bath qubits only differ from the control qubits by static imperfections. Hence, we set the average value of the distribution to $B_{0}^{\alpha}={\cal{B}}^{\alpha}$, and the detuning to $\delta=0.4\ {\epsilon}$. Our simulations included a number of two-qubit interactions. Here we only report our results for $J_{x}= 0.50, \: 1.00, \: 2.00$, with units in ${\epsilon}$, for which we confirmed the onset of chaos in previous studies \cite{CW1,CW2,CW3}.

\subsection{Exact numerical approach}

Exploiting the low temperature limit for the bath density, the time evolved subsystem reduced density can be expressed exactly as 
\begin{equation}
\hat{\rho}_{S} (t)=\sum_{n=1}^{n_{eig}} p_{n} {\rm Tr}_{B} 
[ |\Psi_{n}(t) \rangle \langle \Psi_{n}(t)| ],
\label{TDen2}
\end{equation}
\noindent
where the populations of the bath are given by
\begin{equation}
p_{n}=\frac{e^{-E_{n}/k_BT}}{ \sum_{m=1}^{n_{eig}} e^{-E_{m}/k_BT}},
\end{equation}
\noindent
and where the states $|\Psi_{n}(t) \rangle$ evolve from the Schr\"{o}dinger equation with the total Hamiltonian (\ref{HamQC}) and initial conditions $|\Psi_n(0)\rangle=|\psi_0\rangle\otimes |n\rangle$. In all calculations $n_{eig}=20$ was sufficient for the given low temperature. We used a Lanczos algorithm\cite{Arp} for exact diagonalization of the bath Hamiltonian and an eight-order variable stepsize Runge-Kutta code\cite{RK} for the numerical integrations. 
  
\subsection{Chaotic Kraus decomposition approach}

Analytic solutions for the chaotic Kraus decomposition should be readily obtainable due to the low dimension of the CNOT system. 
However, we employ an alternative approach here by employing numerical solutions of the Schr\"{o}dinger equation.
This is more convenient due the number of initial states, couplings, and observables. Moreover, this approach is generally applicable
for any subsystem. Since the initial states are all pure we used Eq. (\ref{SEs}) and the sum (\ref{SumS}).

We used the $n_{eig}=20$ low lying exact bath eigenstates $|n\rangle$ for $J_{x}=0.50, 1.00, 2.00$ to calculate the $B^{n,n}$ required by the CKD.
For $xx$ and $zz$ type coupling operators $B^{n,n}=\Sigma_{\alpha}^{n,n}= \langle n | \hat{\Sigma}_{\alpha}|n\rangle$ where $\alpha\in \{x,z\}$ stands for bit-flip and phase type couplings, respectively. 

The CKD requires exact bath eigenstates in the calculations of the $B_{\mu}^{n,n}$ terms. Exact diagonalization of the bath Hamiltonian should be easy to achieve for quite large bath dimensions by standard Lanczos matrix diagonalization routines\cite{Arp}, or more
recent generalizations\cite{Tuck}. In cases where the bath dimension is too large, e.g. a large coupled oscillator bath where exact diagonalization is impossible, alternative approaches may be taken to calculate the $B_{\mu}^{n,n}$ and $E_n$. For example, hybrid quantum-semi-classical molecular dynamics simulations can prove very useful to calculate $B_{\mu}^{n,n}$ terms. Perhaps even the Wigner method\cite{GB} would
suffice to calculate $B_{\mu}^{n,n}$ once the $E_n$ are known. These approaches will be investigated elsewhere. 

\section{Results for average purity and fidelity}
 
\begin{figure}
\centering
\includegraphics[width=3in,height=3in]{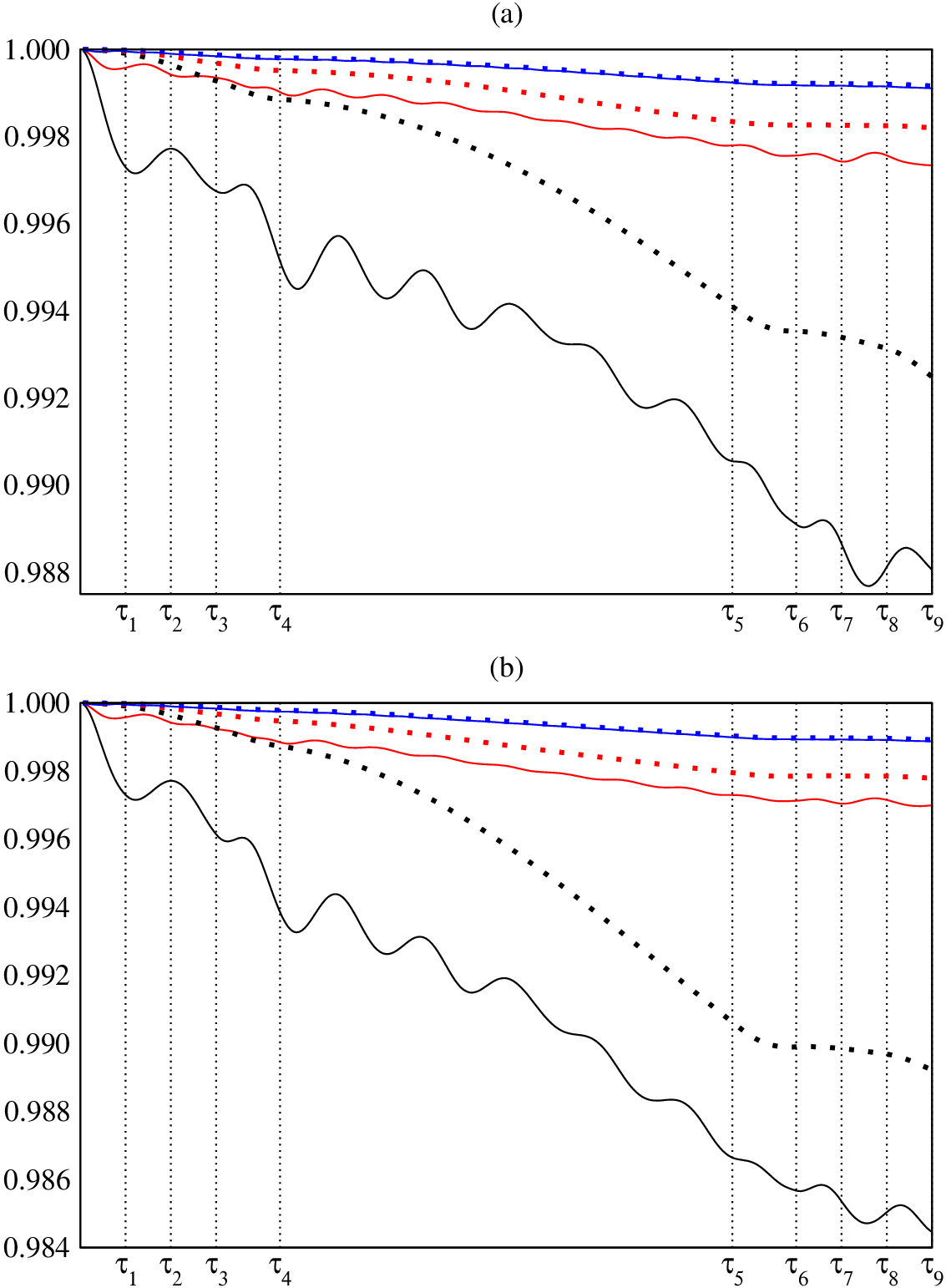}
\caption{Exact numerical (solid lines) and Kraus (dotted lines) results for averaged purity ${\cal P}(t)$ in case of  $xx$-type coupling for $J_{x}=0.50$ (black), $J_{x}=1.00$ (red), and $J_{x}=2.00$ (blue). (a) Standard basis states and (b) Bell states.}
\label{purxx}
\end{figure}

\begin{figure}
\centering
\includegraphics[width=3in,height=3in]{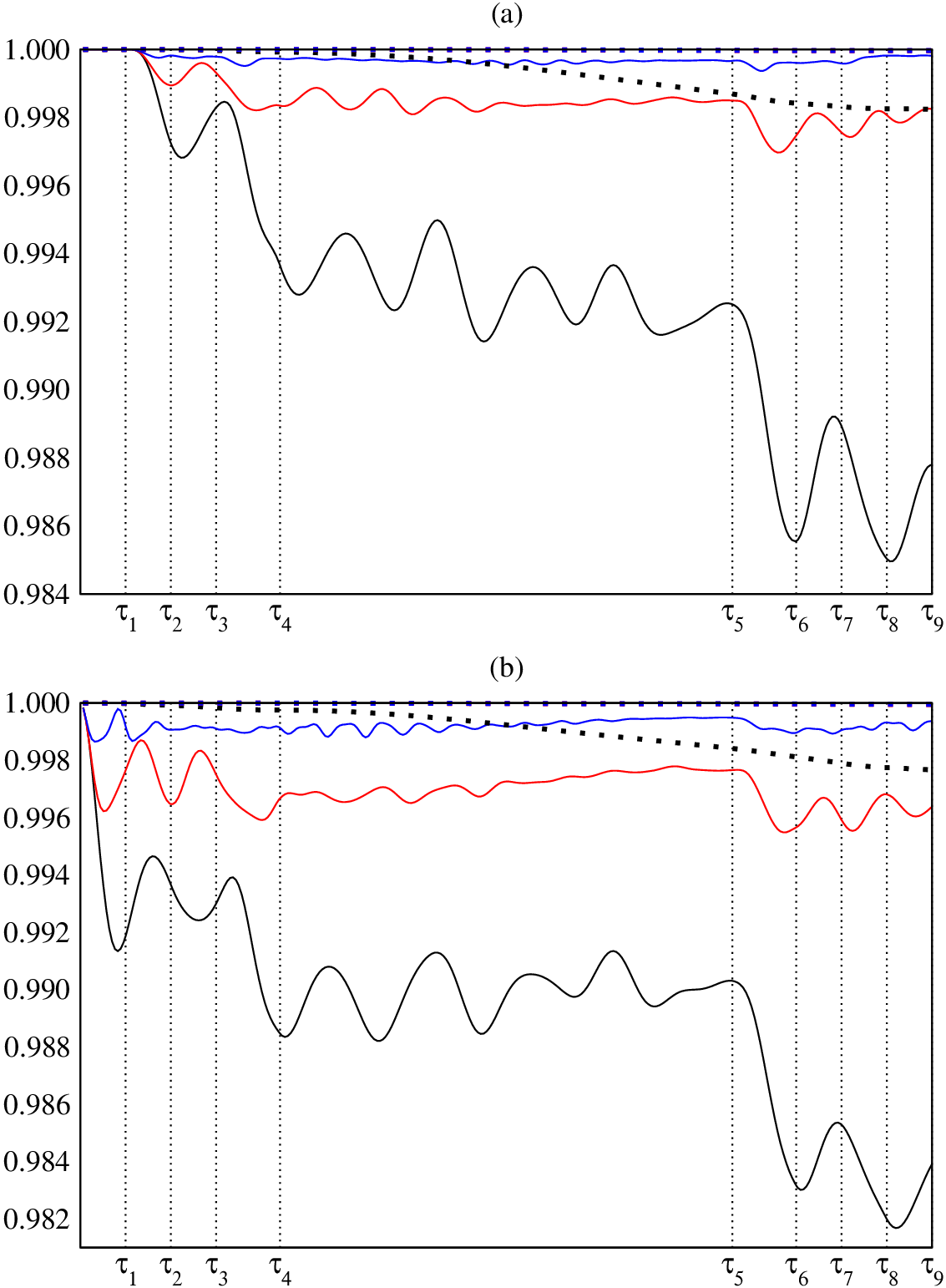}
\caption{Exact numerical (solid lines) and Kraus (dotted lines) results for averaged purity ${\cal P}(t)$ in case of  $zz$-type coupling for $J_{x}=0.50$ (black), $J_{x}=1.00$ (red), and $J_{x}=2.00$ (blue). (a) Standard basis states and (b) Bell states.}
\label{purzz}
\end{figure}

We plot average purity $\bar{\cal P}(t)$ for $xx$-type bit-flip coupling in Figure \ref{purxx} and for $zz$-type phase coupling in Figure \ref{purzz} for three different values of intra-bath coupling, $J_x=0.50, 1.00, 2.00$. Decoherence and dissipation result
in a purity decay of less than 1 \% over the course of the gate. The exact numerical results are represented by solid lines and Kraus results are represented by dotted lines, and each coupling value $J_x$ is assigned to the same color/line convention throughout. Results for standard basis states and Bell states are shown in subfigures (a) and (b), respectively. Switching times of elementary gate operations are also indicated by the grid lines.  

Figure \ref{purxx} (a) shows excellent quantitative agreement between the exact and CKD for the most chaotic case of $J_x=2$. 
For $J_x=1$ the agreement is also quite good with errors in the purity of less than .05 \%. For the least chaotic case of $J_x=.5$ the discrepancy is on the order of .2 \%. On the whole 
these results are surprisingly accurate for such a small bath. The results for Bell states in Figure \ref{purxx} (b) are virtually indistinguishable
from those of (a).

Figure \ref{purzz} (a) is quite different from Figure \ref{purxx} (a). The overall exact decay of purity is comparable, but the CKD discrepancies for $J_x=2$ are on the order of .1 \%, while those for 
$J_x=1$ are about .2 \%. For $J_x=.5$ the error is terrible. Figure \ref{purzz} (b) for Bell states is again the same as (a) for standard states. 

\begin{figure}
\centering
\includegraphics[width=3in,height=3in]{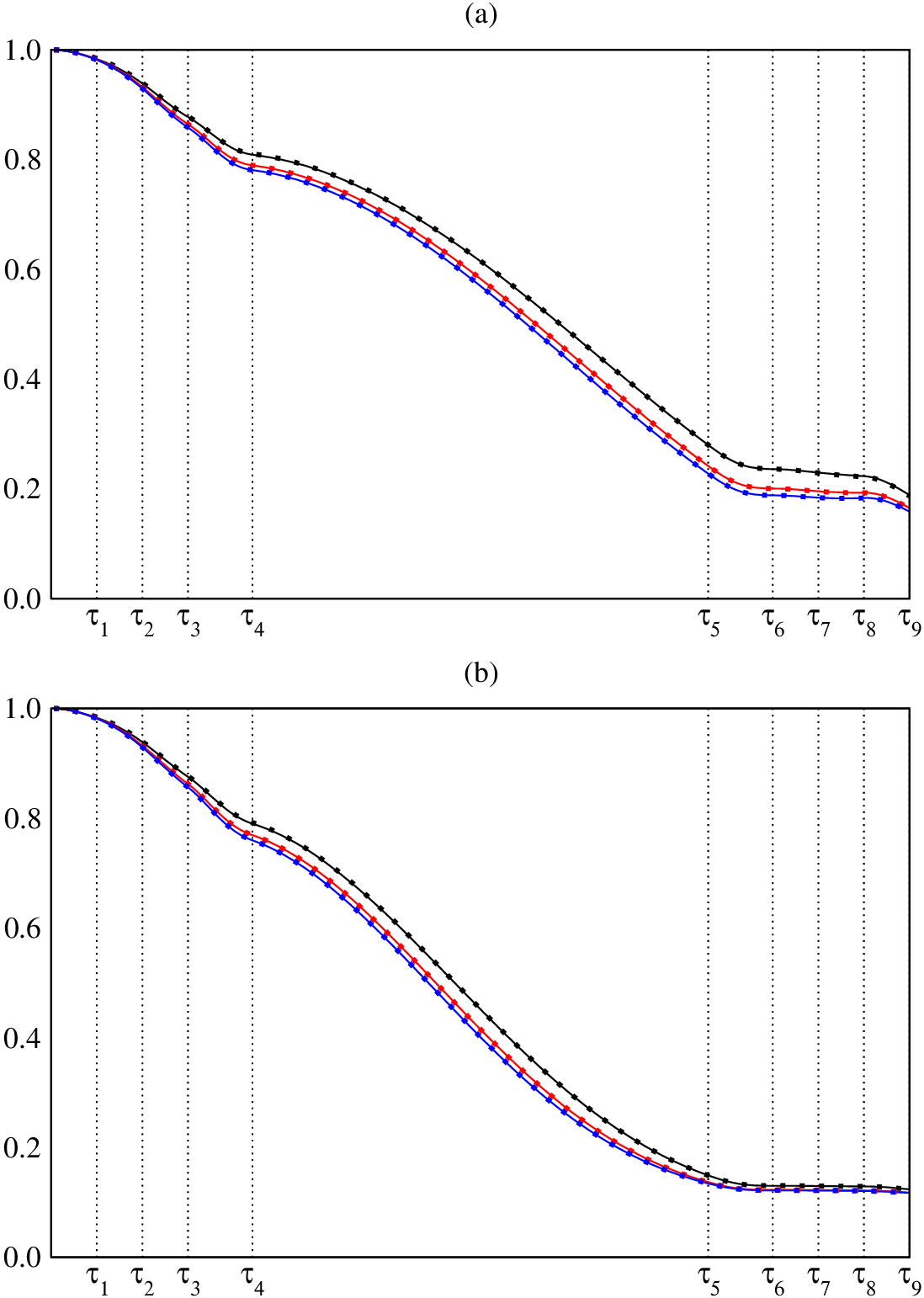}
\caption{Exact numerical (solid lines) and Kraus (dotted lines) results for averaged fidelity ${\cal F}(t)$ in case of $xx$-type coupling for $J_{x}=0.50$ (black), $J_{x}=1.00$ (red), and $J_{x}=2.00$ (blue). (a) Standard basis states and (b) Bell states.}
\label{fidxx}
\end{figure}

\begin{figure}
\centering
\includegraphics[width=3in,height=3in]{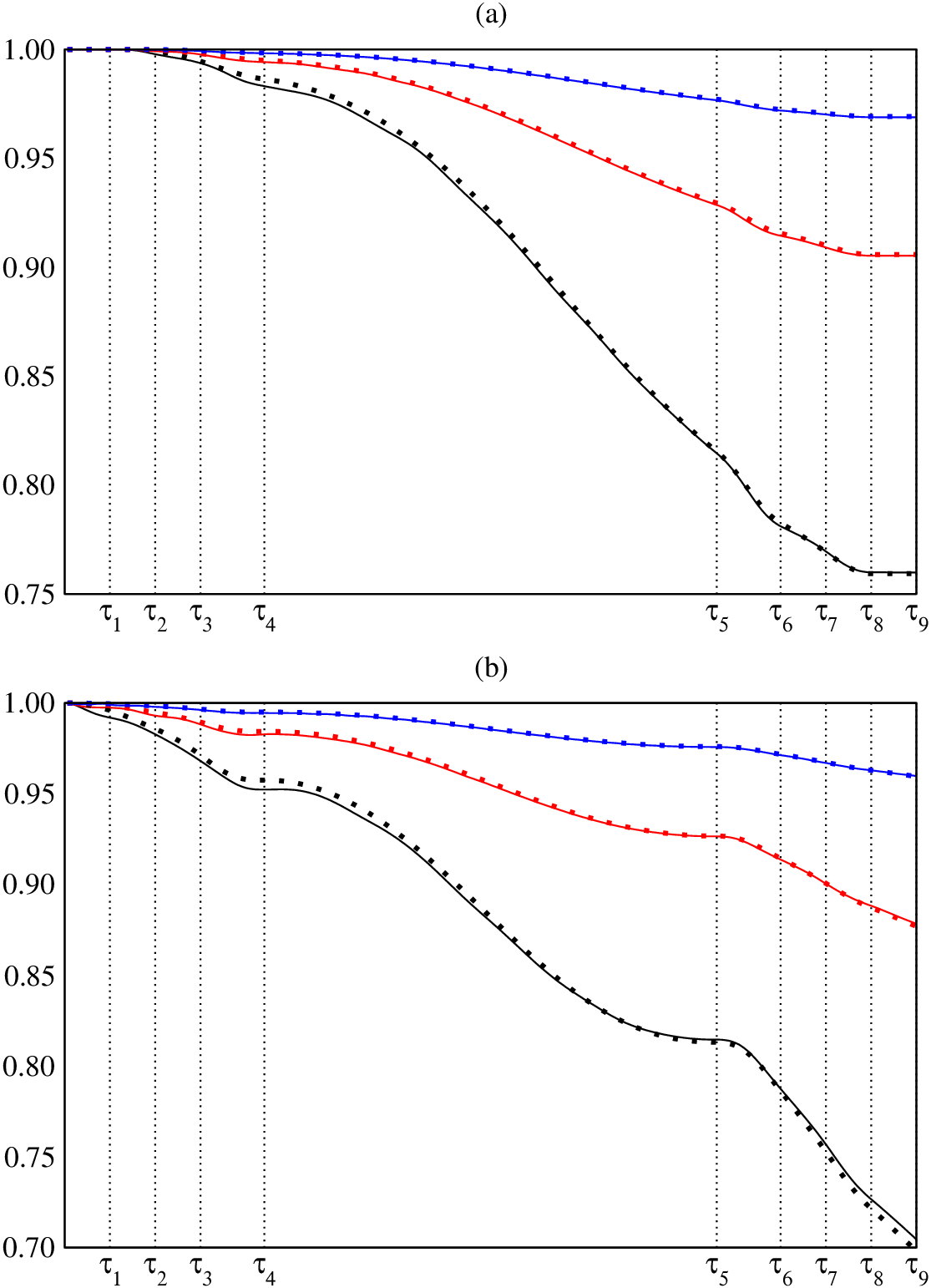}
\caption{Exact numerical (solid lines) and Kraus (dotted lines) results for averaged fidelity ${\cal F}(t)$ in case of  $zz$-type coupling for $J_{x}=0.50$ (black), $J_{x}=1.00$ (red), and $J_{x}=2.00$ (blue). (a) Standard basis states and (b) Bell states.}
\label{fidzz}
\end{figure}

We plot $\bar{\cal F}(t)$ for $xx$-type coupling in Figure \ref{fidxx} and for $zz$-type coupling in Figure \ref{fidzz}, for the same $J_{x}$ values. Here,
we see perfect agreement between exact and CKD predictions for all states, all $J_x$ values, and all couplings. This is all the more remarkable when
one notes that these errors are now very large. The fidelity in Figure \ref{fidxx} decays to 10 \% of its initial value, while that in 
Figure \ref{fidzz} decays to 70 \% of its initial value. The average fidelity of Figure \ref{fidxx} for $xx$ coupling shows no sensitivity to $J_x$, while that in Figure \ref{fidzz} for $zz$ coupling varies substantially with $J_x$. The CKD captures both these effects. The large magnitude of the infidelity in both cases is due to a coherent shift of 
the subsystem\cite{CW1,CW2,CW4}. Detailed discussions of the reasons for the different 
behaviors for the different coupling are given in Ref. \cite{CW4}. 

\section{Results for matrix elements of reduced density}

Purity and fidelity suffice for an overall identification of the magnitudes of non-unitary and unitary errors.
The comparison of ideal and actual reduced density matrix elements, however, provides further valuable information 
on what actually goes wrong in an algorithm during open system dynamics. Here again it is impossible for us to
present all of this data. As a generic representation of our results, we present a comparison of these matrix elements 
for two initial subsystem states; $|11\rangle$ is representative of the standard basis states, and $(|00\rangle+|11\rangle)/\sqrt{2}$ is 
representative of Bell states. We consider both $xx$-type and $zz$-type couplings for $J_{x}=1.00 \epsilon$. 

\begin{figure}
\centering
\includegraphics[width=3in,height=3.5in]{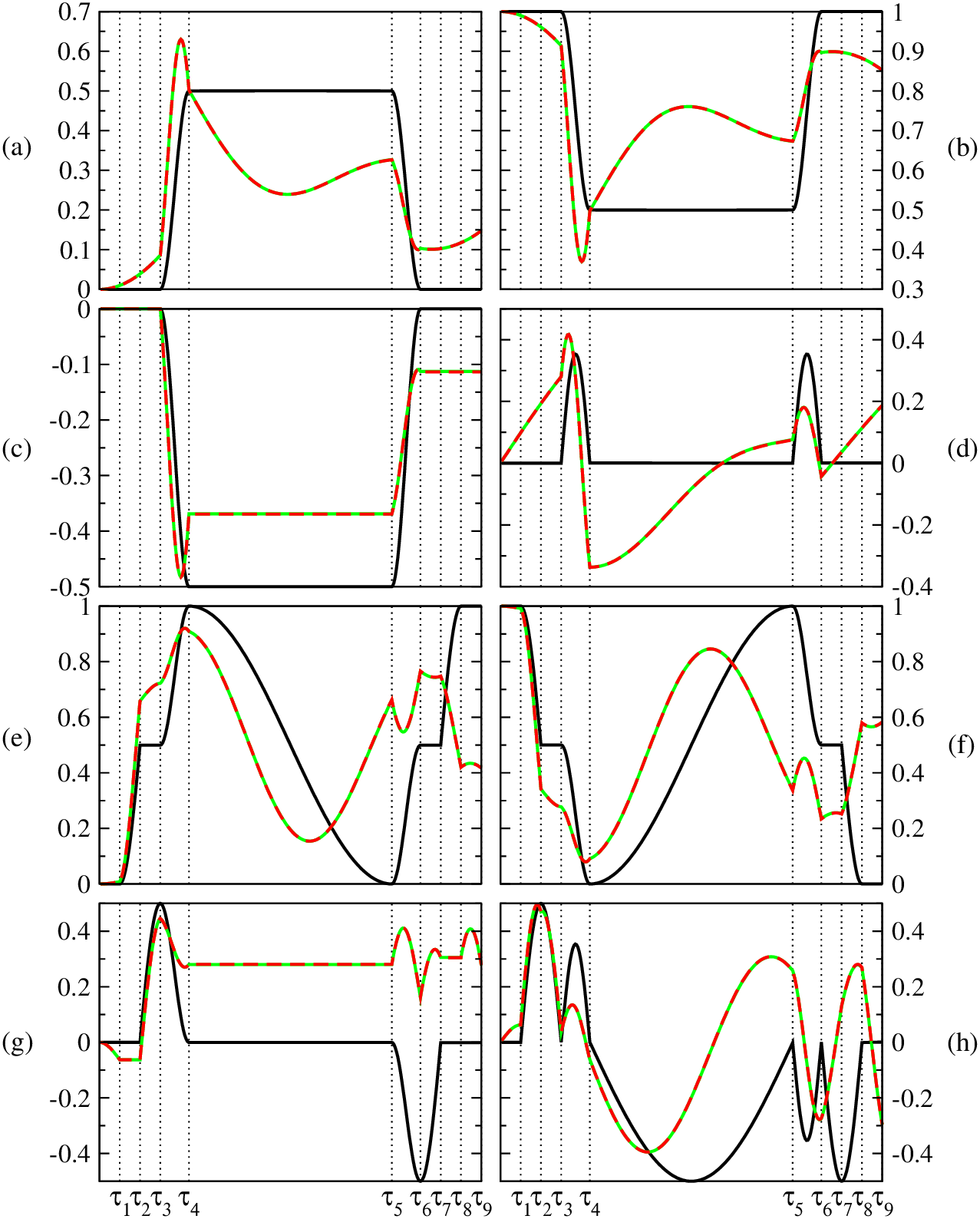}
\caption{Matrix elements of reduced density of first and second qubits for $xx$-type  coupling. The initial state of system is $|11\rangle$ and the intra-bath coupling $J_{x}=1.00\ \epsilon$.} 
\label{f4xx}
\end{figure}

\begin{figure}
\centering
\includegraphics[width=3in,height=3.5in]{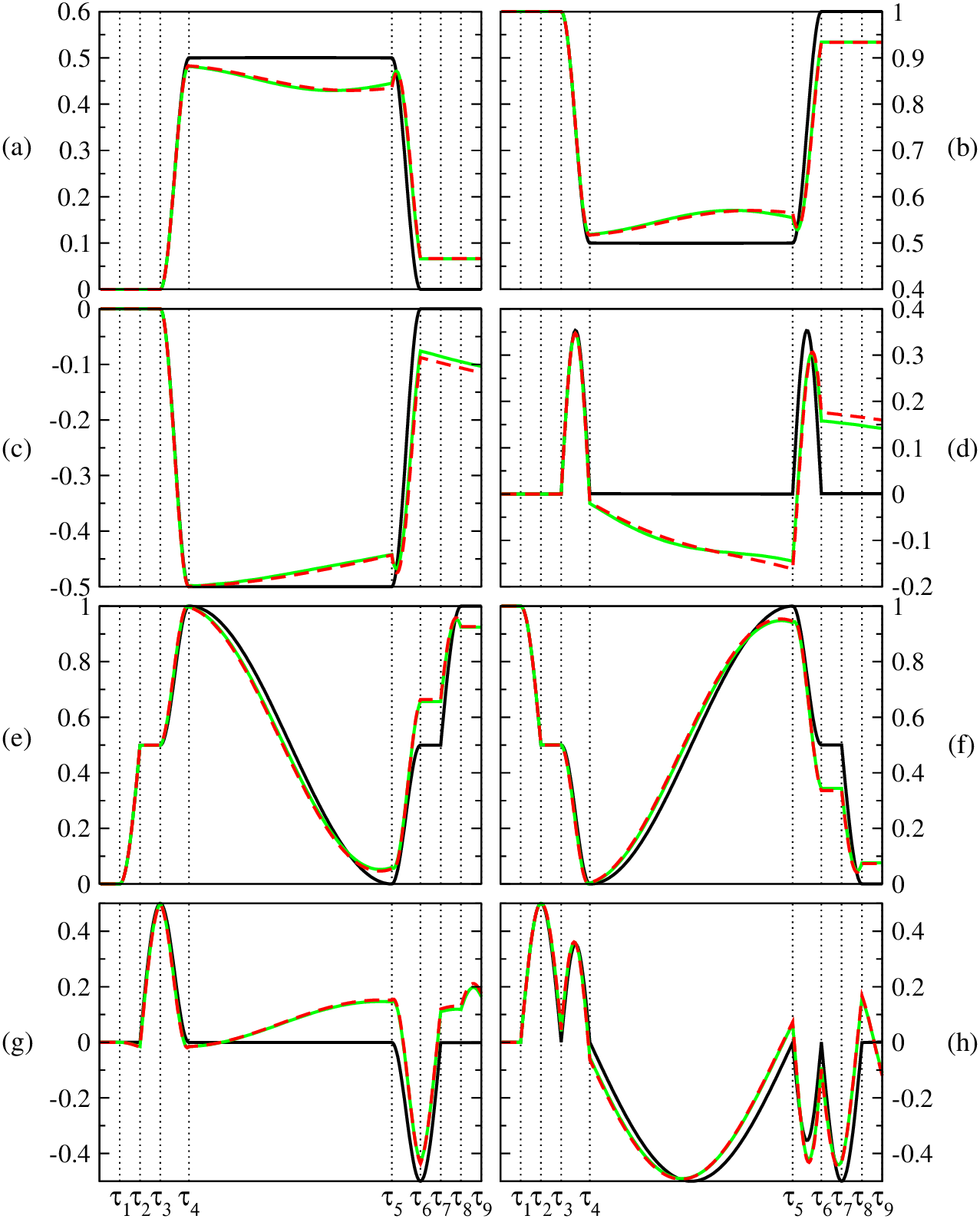}
\caption{Matrix elements of reduced density of first and second qubits for $zz$-type  coupling. The initial state of system is $|11\rangle$ and the intra-bath coupling $J_{x}=1.00\ \epsilon$.} 
\label{f4zz}
\end{figure}

We compare the matrix elements for $xx$-type coupling in Figure \ref{f4xx} and $zz$-type coupling in Figure \ref{f4zz} for the initial state $|11\rangle$. 
In these and subsequent figures the error-free coherent time evolution is given by black solid lines. Exact open dynamics time evolutions are denoted by solid green lines, and the CKD results are given by dashed red lines. 

Each subfigure in a figure represents a different matrix element. The specific matrix element plotted in each subfigure is as follows: in subfigure (a) we plot
${\hat{\rho}}_{00}^{(1)}(t)=\langle 0|{\rm Tr}_2[{\hat{\rho}}_{S}(t)]|0\rangle$. Similarly, in subfigure (b) we plot
${\hat{\rho}}_{11}^{(1)}(t)=\langle 1|{\rm Tr}_2[{\hat{\rho}}_{S}(t)]|1\rangle$, 
in subfigure (c)
${\rm Re}\{\hat{\rho}_{01}^{(1)}(t)\}={\rm Re}\{\langle 0|{\rm Tr}_2[{\hat{\rho}}_{S}(t)]|1\rangle\}$ and in subfigure (d)
${\rm Im}\{\hat{\rho}_{01}^{(1)}(t)\}={\rm Im}\{\langle 0|{\rm Tr}_2[{\hat{\rho}}_{S}(t)]|1\rangle\}$. 
Similarly, 
in subfigure (e) we plot
${\hat{\rho}}_{00}^{(2)}(t)=\langle 0|{\rm Tr}_1[{\hat{\rho}}_{S}(t)]|0\rangle$, 
in subfigure (f)
${\hat{\rho}}_{11}^{(2)}(t)=\langle 1|{\rm Tr}_1[{\hat{\rho}}_{S}(t)]|1\rangle$, 
in subfigure (g)
${\rm Re}\{\hat{\rho}_{01}^{(2)}(t)\}={\rm Re}\{\langle 0|{\rm Tr}_1[{\hat{\rho}}_{S}(t)]|1\rangle\}$, and in subfigure (h)
${\rm Im}\{\hat{\rho}_{01}^{(2)}(t)\}={\rm Im}\{\langle 0|{\rm Tr}_1[{\hat{\rho}}_{S}(t)]|1\rangle\}$. 

The agreement between the exact and CKD results is excellent in all cases. The deviations from ideal free evolution are large in all cases, while
the discrepancies between the exact and CKD results are basically negligible. The worst deviations are again seen in the $zz$ coupling case but
these are still very small.

\begin{figure}
\centering
\includegraphics[width=3in,height=3.5in]{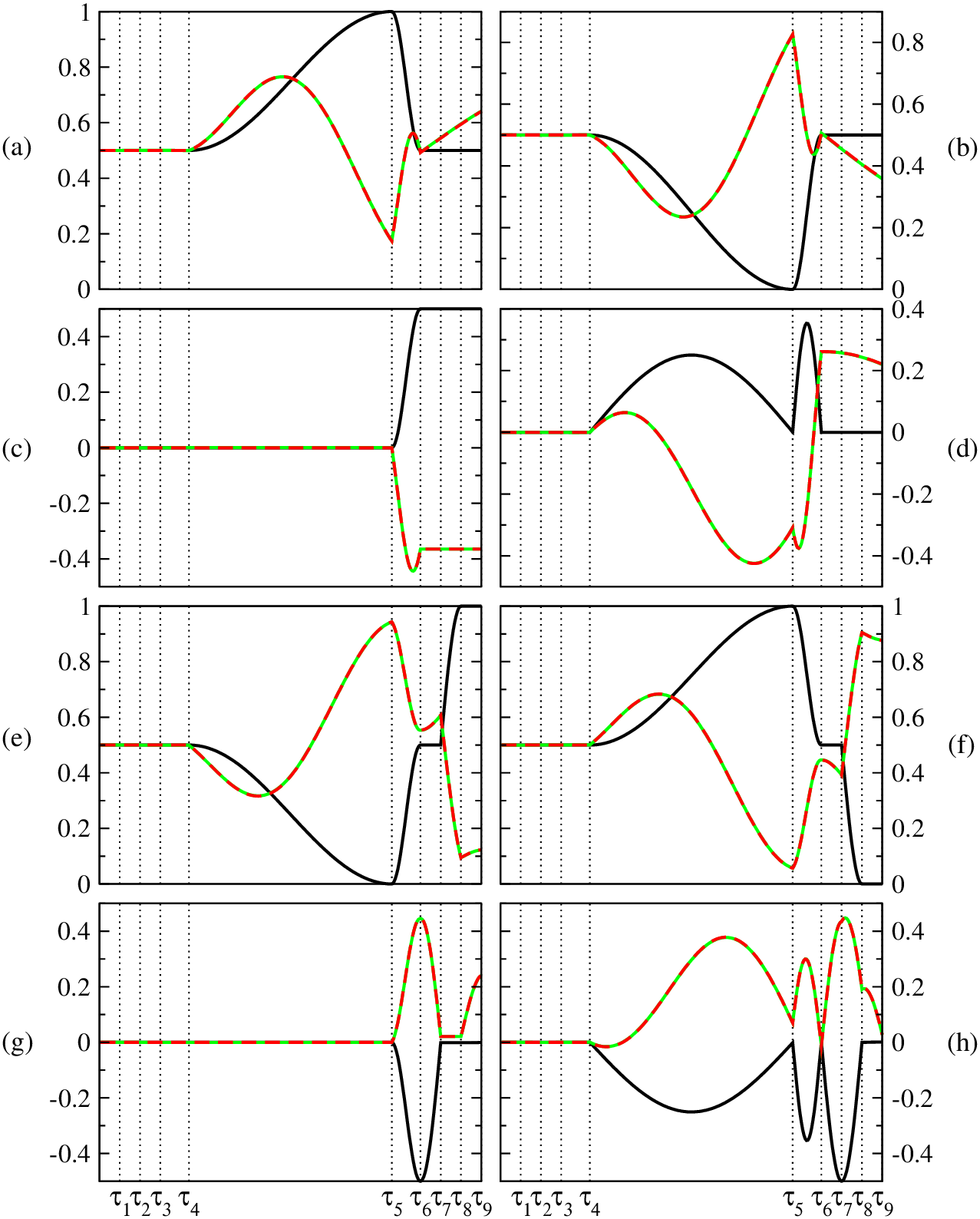}
\caption{Matrix elements of reduced density of first and second qubits for $xx$-type  coupling. The initial state of system is a Bell state of the form $(|00\rangle+|11\rangle)/\sqrt{2}$ and the intra-bath coupling $J_{x}=1.00\ \epsilon$.}
\label{b1xx}
\end{figure}

\begin{figure}
\centering
\includegraphics[width=3in,height=3.5in]{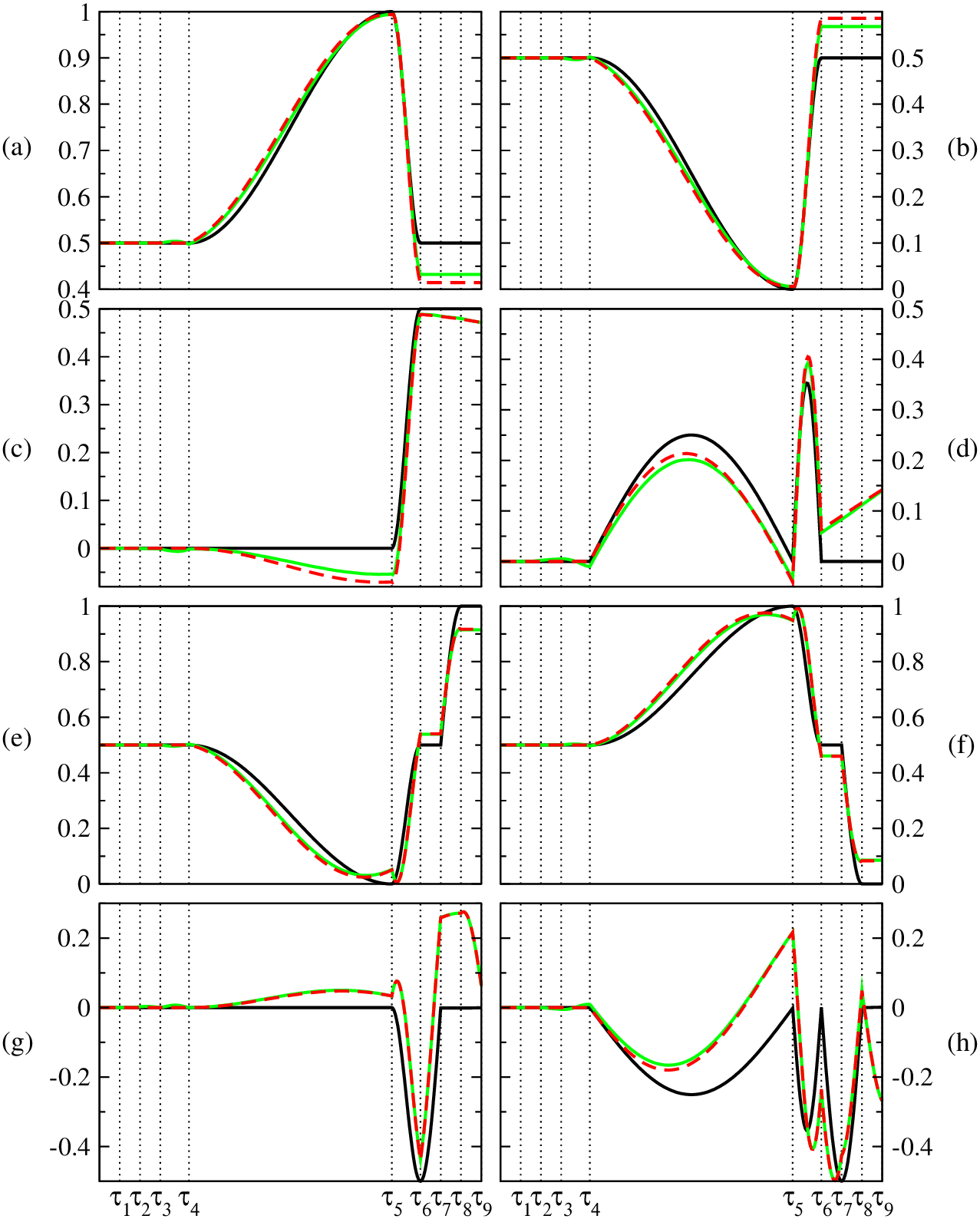}
\caption{Matrix elements of reduced density of first and second qubits for $zz$-type  coupling. The initial state of system is a Bell state of the form $(|00\rangle+|11\rangle)/\sqrt{2}$ and the intra-bath coupling $J_{x}=1.00\ \epsilon$.}
\label{b1zz}
\end{figure}

We show results for the matrix elements for the initial Bell state $(|00\rangle+|11\rangle)/\sqrt{2}$ in Figure \ref{b1xx} for $xx$-type coupling and in Figure \ref{b1zz} for $zz$-type coupling.  Here again the agreement between the exact and CKD predictions is astonishingly good. The only visible deviations occur for $zz$ coupling. See Figure \ref{b1zz} (a) and (c), for example, where there are some small deviations.

\section{Summary and discussion}

We have considered a quantum system interacting simultaneously with time-dependent external driving fields and chaotic baths of thermodynamic dimension. We derived a Kraus decomposition for the subsystem reduced density in an explicit and computationally tractable form which is exact for baths of 
thermodynamic dimension. We tested the accuracy of the Kraus decomposition against exact numerical results for a model system consisting of an isolated statically flawed QC performing a CNOT gate. We obtained quite accurate results for a large number of different configurations in spite of our small bath dimension. These promising results suggest that the Kraus decomposition can be a very useful and practical computational tool for low temperature simulations of open quantum systems. In particular, the CKD could prove quite useful for determining low temperature chemical reaction rates via the half-collision problem.

\ack{The authors gratefully acknowledge the support of the Natural Sciences and Engineering Research Council of Canada and computing resources provided by WestGrid.}


\begin{thebibliography}{99}

\bibitem{QC} M.A. Nielsen, I.L. Chuang, Quantum Computation and Quantum Information, Cambridge University Press, Cambridge, 2000.

\bibitem{GS} B. Georgeot and D.L. Shepelyansky, Phys. Rev. E 62 (2000) 3504; B. Georgeot and D.L. Shepelyansky, Phys. Rev. E 62 (2000) 6366; G. Benenti, G. Casati, and D.L. Shepelyansky, Eur. Phys. J. D 17 (2001) 265. 

\bibitem{CW1} M. \c{C}etinba\c{s} and J. Wilkie, Probing internal bath dynamics by a Rabi oscillator-based detector, Phys. Lett. A (2007), doi:10.1016/j.physleta.2007.05.075, in press; arXiv: 0705.4018.   

\bibitem{CW2} M. \c{C}etinba\c{s} and J. Wilkie, Quantum pathology of static internal imperfections in flawed quantum computers, Phys. Lett. A (2007), doi:10.1016/j.physleta.2007.05.074, in press; arXiv: 0705.4017.  

\bibitem{CW4} M. \c{C}etinba\c{s} and J. Wilkie, Manifold algorithmic infidelity in quantum computers with static internal imperfections, manuscript in preparation.

\bibitem{LEKT} W. Wang and B. Li, Phys. Rev. E 66, 056208 (2002); Ph. Jacquod, P.G. Silvestrov, and C.W.J. Beenakker, Phys. Rev. E 64, 055203 (2001); R.A. Jalabert and H. M. Pastawski, Phys. Rev. Lett. 86, 2490 (2001). 

\bibitem{Brown} Brown K R, Harrow A W and Chuang I L 2000 Phys. Rev. A {\bf 70} 052318

\bibitem{SB} M. Shapiro and P. Brumer, Principles of the Quantum Control of Molecular Processes, Wiley-Interscience, Hoboken, 2003.

\bibitem{Rat} M. Ratner and D. Ratner, Nanotechnology, Prentice Hall, Upper Saddle River, NJ, 2003.

\bibitem{Brumer} V.S. Batista and P. Brumer, Phys. Rev. Lett. 89 (2002) 143201; J. Gong and P. Brumer, Phys. Rev. A 68 (2003) 022101.

\bibitem{Red} A.G. Redfield, IBM J. Res. Dev., 1, (1957) 19 ; Adv. Magn. Reson., 1, (1965) 1.

\bibitem{Opp} P. Gaspard and M. Nagaoka, J. Chem. Phys., 111 (1999) 5668; 
A. Su\'{a}rez, R. Silbey and I. Oppenheim, J. Chem. Phys., 97 (1992) 5101; 
V. Romero-Rochin and I. Oppenheim, J. Stat. Phys., 53 (1988) 307; 
Physica A, 155 (1989) 52; 
V. Romero-Rochin, A. Orsky and I. Oppenheim, {\em ibid.}, 156 (1989) 244.

\bibitem{Coal} R.D. Coalson and D.G. Evans, Chem. Phys. 296 (2004) 117.

\bibitem{Neu} A.A. Neufeld, J. Chem. Phys. 119 (2003) 2488.

\bibitem{Bulg} A. Bulgac, G.D. Dang and D. Kusnezov, Phys. Rev. E 58 (1998) 196.

\bibitem{GB}
J. Gong and P. Brumer, Phys. Rev. Lett. 90 (2003) 050402;
J. Gong and P. Brumer, J. Mod. Opt. 50 (2003) 2411.

\bibitem{SRA1} J. Wilkie, Phys. Rev. E 62  (2000) 8808.

\bibitem{SRA2} J. Wilkie, J. Chem. Phys. 114 (2001) 7736

\bibitem{SRA3} J. Wilkie, J. Chem. Phys. 115 (2001) 10335.

\bibitem{CW5} M. \c{C}etinba\c{s} and J. Wilkie, Mean field master equation for self-interacting environments,
manuscripts in preparation.

\bibitem{dsg} Lindblad G 1976 {\it Commun. Math. Phys.} {\bf 48} 119; \\
Gorini V, Kossakowski A, and Sudarshan E C G 1976 {\it J. Math. Phys.} 
{\bf 17} 821; \\
Alicki R and Lendi K 1987 {\it Quantum Dynamical Semigroups and
Applications} (Berlin: Springer)

\bibitem{Kraus} K. Kraus, Ann. Phys. (N.Y.), 64 (1971) 311; 
K. Kraus, States, Effects and Operations: Fundamental Notions of Quantum Theory, Springer-Verlag, Berlin,1983.

\bibitem{CW3} M. \c{C}etinba\c{s} and J. Wilkie, Kraus decomposition for chaotic environments, Phys. Lett. A (2007), doi:10.1016/j.physleta.2007.08.064, in press; arXiv: 0709.3608.  

\bibitem{Abs} See for example Y. Sajeev, M. Sindelka and N. Moiseyev, Chem. Phys. 329, 307 (2006) and references therein.

\bibitem{Copt} H. Hayashi, G. Kimura, and Y. Ota Phys. Rev. A 67 (2003) 062109; 
P. \v{S}telmachovi\v{c} and V. Bu\v{z}ek Phys. Rev. A 64 (2001) 062106; 67 (2003) 029902(E).

\bibitem{SemiC} J. Wilkie and P. Brumer, Phys. Rev. A 55 (1997) 43; T. Prosen, Ann. Phys. 235 (1994) 115; M. Feingold and A. Peres, Phys. Rev. A 34 (1986) 591; P. Pechukas, Phys. Rev. Lett. 51 (1983) 943.

\bibitem{Magnus} W. Magnus, Commun. Pure Appl. Math. 7 (1954) 649.  

\bibitem{RK} E. Hairer, S.P. Norsett, and G. Wanner, Solving Ordinary Differential Equations I. Nonstiff Problems, 2nd Ed., Springer Series in Computational Mathematics, Vol. 8, Springer-Verlag, Berlin; New York, 1993. 

\bibitem{Chaos} F. Haake, Quantum Signatures of Chaos, 2nd edn, Springer, Berlin, 2001.

\bibitem{Chaos2} {\em Chaos and Quantum Physics}, edited by M.--J. Giannoni, A. Voros, and J. Zinn--Justin, Les Houches Lecture Series Vol. 52, North--Holland, Amsterdam, 1991.

\bibitem{Peres} A. Peres, Phys. Rev. A 30 (1984) 1610.

\bibitem{Emerson} see for example J. Emerson, Y.S. Weinstein, S. Lloyd and D.G. Cory, Phys. Rev. Lett. 89 (2002) 284102 and references therein. 

\bibitem{Nori} You, J. Q. and Tsai, J. S. and Nori, Franco, Phys. Rev. Lett. 89 (2002) 197902.

\bibitem{Makhlin} Y. Makhlin, G. Sch\"{o}n,~and A. Shnirman, Rev. Mod. Phys., 73 (2001) 357 and references therein.

\bibitem{Arp} R.B. Lehoucq, D.C. Sorensen, and C. Yang, ARPACK Users' Guide: Solution of Large-Scale Eigenvalue Problems with Implicitly Restarted Arnoldi Methods, SIAM, Philadelphia, 1998.

\bibitem{Tuck} R. Dawes and T. Carrington, J. Chem. Phys. 124, 054102 (2006); {\em ibid} 122, 134101 (2005).




\end{thebibliography}
\end{document}